\documentclass[10pt,prb,twocolumn,showpacs]{revtex4}
\usepackage{graphicx}
\usepackage{epsf}
\usepackage{ifthen}
\usepackage{amsmath}
\usepackage{amssymb}
\usepackage{subfigure}
\usepackage{psfrag}
\usepackage{float}
\usepackage{subeqnarray}
\usepackage{color}
\usepackage{booktabs}
\begin{document}

\title{Stable Half-Metallic Monolayers of FeCl$_{2}$}

\author{E. Torun}
\email{engin.torun@uantwerpen.be}
\affiliation{Department of Physics, University of Antwerp, 2610,Antwerp, Belgium}

\author{H. Sahin}
\affiliation{Department of Physics, University of Antwerp, 2610,Antwerp, Belgium}

\author{S. K. Singh}
\altaffiliation{Current address: Department of Science and Technology, Link\"{o}ping University, 60174 Norrk\"{o}ping, Sweden}

\author{F. M. Peeters}
\affiliation{Department of Physics, University of Antwerp, 2610,Antwerp, Belgium}

\date{\today}


\begin{abstract}

The structural, electronic and magnetic properties of single layers of Iron 
Dichloride (FeCl$_{2}$) were calculated using first principles calculations. We 
found that the 1T phase of the single layer FeCl$_{2}$ is 0.17 eV/unit cell more favorable 
than its 1H phase. The structural stability is confirmed by phonon 
calculations. We found that 1T-FeCl$_{2}$ possess three Raman-active (130, 179 
and 237 cm$^{-1}$) and one Infrared-active (279 cm$^{-1}$) phonon branches. 
The electronic band dispersion of the 1T-FeCl$_{2}$ is calculated using both GGA-PBE and 
DFT-HSE06 functionals. Both functionals reveal that the 1T-FeCl$_{2}$ has a 
half-metallic ground state with a Curie temperature of 17 K.
\end{abstract}
\maketitle
The synthesis of stable two-dimensional (2D) materials \cite{Novoselov,Najmaei2013} and observing their various exceptional physical properties triggered a large interest in their potential  technological applications. Single layer transition-metal dichalcogenides (TMDs) are a new class of materials which are promising candidates for next generation of flexible nanoelectronic devices due to their  wide range of important properties in their bulk and monolayer form such as superconductivity\cite{sipos,takada} and  half-metallicity. \cite{shishidou,jin,antonov,saha,leal}

The single layer TMDs can be obtained from their three-dimensional (3D) counterpart which consist of weakly interacting 2D layered structures. Most of these materials have either D$_{6f}$ (2H structure) or D$_{3d}$ (1T structure) point-group symmetry and very few of them are stable in both the 1T and 2H phases. A lot of efforts are being done to predict new 2D materials theoretically and synthesize them experimentally with tailored electronic, optical, and chemical properties. 

Half-metallicity is one of the desired properties for future 
spintronics devices and it is observed in 3D compounds several decades ago. \cite{groot} Most of the 2D materials do not have intrinsic 
half-metallicity but it has been reported that graphene~\cite{son2007, Hod2007} 
and hydrogenated boron-nitride nanoribbons~\cite{samarakoon2012} gain 
half-metallicity when in-plane electric fields were applied. Chemical 
modification of the edges with different  functional groups is also an 
alternative approach to achieve half-metallicity in 2D 
structures.\cite{Martins2008, Katsnelson2008}  It has been recently reported 
that graphitic carbon nitride (g-C$_4$N$_3$)\cite{hashmi,hu}, MnO$_2$ nanosheet 
with vacancies and monolayer chromium nitride \cite{zhang} show half-metallic 
behaviour.  

\begin{figure}
\includegraphics[width=7.5cm]{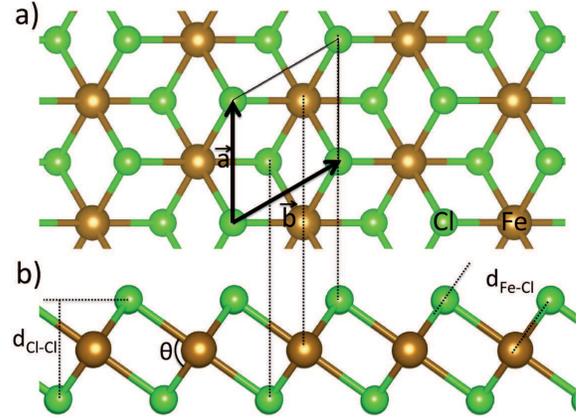}
\caption{\label{fig1}
(Color online) Atomic structure of 2D single-layer FeCl$_2$ in the T structure.
(a) Top and (b) side views of the T structure showing the primitive unit cell
of the 2D hexagonal lattice with Bravais lattice vectors $\vec{a}$ and $\vec{b}$
($|\vec{a}|$=$|\vec{b}|$ and relevant internal structural parameters.}
\end{figure}

FeCl$_2$, which is a metal in its bulk form, is another member of family of TM crystals.
Its bulk and molecular form have been investigated. Saraireh and Altarawneh ~\cite{Saraireh} investigated the structure and thermodynamics stability of the FeCl$_2$ surface using DFT calculations and found that the FeCl$_2$ (100-Cl) surface is the most stable configuration. The electronic energy levels and the M\"{o}ssbauer spectroscopic parameters of FeCl$_2$ were calculated by Bominaar et al.~\cite{Bominaar} and by Mishra et al.~\cite{Mishra} at the self-consistent field level. The X-ray properties of FeCl$_2$ were reported by  Chou et al.~\cite{Chou} and Veal et al.~\cite{Veal} using density functional theory. Shechter et al. performed a M\"{o}ssbauer resonance study on  FeCl$_2$ monolayers deposited on oriented basal planes of graphite. \cite{shechter}  

In this work we will investigate the structural, electronic and vibrational properties of monolayer FeCl$_2$ by using first principle calculations.  We will show that the minimum energy configuration of the single layer FeCl$_2$ is the T structure which has a half-metallic behaviour. 

All the calculations are performed using the projector augmented wave  (PAW)~\cite{blochl1994} potentials as implemented in the Vienna Ab-initio Simulation Package  (VASP).~\cite{vasp1,vasp4}  The electronic exchange-correlation potential is treated within the spin polarized generalized gradient approximation (GGA) of Perdew-Burke-Ernzerhof (PBE).~\cite{GGA-PBE1}. A plane-wave basis set with kinetic energy cutoff of 500 eV is used. A vacuum spacing of more  than 15 {\AA} is taken to prevent layer-layer interactions. A set of (20$\times$20$\times$1) $\Gamma$ centered \textbf{k}-point sampling is used for the  primitive unit cell and scaled according to the sizes of the supercells used. The structures are relaxed until self-consistency is reached at the level of 10$^{-5}$ eV between two consecutive steps and the atomic Hellman-Feynman force is  less than 10$^{-4}$ eV/{\AA}. Pressures on the lattice unit cell are decreased to values less than 1.0 kBar. For the charge transfer analysis, the effective charge on atoms is obtained by  the Bader method.~\cite{Henkelman} More accurate electronic structure calculations were performed using the screened-nonlocal-exchange Heyd-Scuseria-Ernzerhof (HSE) functional of the  generalized Kohn-Sham scheme. \cite{heyd,fuchs}  


Experimental and theoretical studies have shown that most of the compounds of  transition metals form single layer crystal structures in T (TaS$_{2}$, VS$_{2}$ HfSe$_{2}$) or H (MoS$_{2}$, WSe$_{2}$, MoTe$_{2}$) phase. Our calculations  show that the minimum energy configuration of a single layer FeCl$_{2}$ is the 1T structure. 1H structure of FeCl$_2$ is 0.17 eV/unit cell higher in energy  than the 1T structure. In the 1T configuration, the Fe atoms are sandwiched between layers of trigonally arranged  Cl atoms and each Fe atom is coordinated to six Cl atoms. The optimized atomic  structure of hexagonal FeCl$_2$ lattice is shown in Fig.~\ref{fig1} and  corresponding structural parameters are presented in Table~\ref{table1}. The  calculated lattice parameter, 3.47 \AA, is in good agreement with previously reported experimental  and theoretical values which are around 3.5 \AA. \cite{Saraireh, vettier}  However, the optimized lattice parameter of the H phase  is 3.34 \AA~which is smaller than the  one for the T phase. The Fe-Cl bond length is 2.45 \AA~in the T phase and is almost the same in the H phase which is 2.47 \AA. Another possible phase of monolayer TMDs is the distorted T structure which was observed in ReS$_2$. \cite{tongay} In ReS$_2$, expanding the lattice parameter of the 1T-ReS$_2$ in one direction leads to a dimerization of Re atoms. This is responsible for an energetically more favorable distorted T structure. A similar investigation for the FeCl$_2$ monolayer revealed that the distorted T phase is not likely in this compound. This observation fits with earlier theoretical and experimental works. \cite{Saraireh,shechter,vettier,wilkinson} 

The cohesive energy per unit cell of FeCl$_2$  is calculated by subtracting the  
total energies of the individual atoms from the total energy of FeCl$_2$ which 
is  formulated as $E_c= E_T[Fe]+2E_T[Cl]-E_T[{FeCl_2}]$. We found that the 
cohesive  energy of FeCl$_{2}$ per unit cell is 2.64 eV/atom in the T phase 
which indicates a strong  cohesion relative to the constituent free atoms and it 
is slightly higher than the value for the H structure which is 2.59 eV/atom. 

\begin{table}[htbp]
\caption{\label{table1} Calculated values of stable, free-standing, 2D single-layer
FeCl$_2$ in the T structure: lattice constant, $|\vec{a}|$=$|\vec{b}|$; bond lengths,
d$_{Fe-Cl}$ and d$_{Cl-Cl}$; Cl-Fe-Cl bond angle, $\theta$; total magnetic moment per
the unit cell, $\mu$; donated (receieved) electron by per Fe (Cl) atoms, $\rho_{Fe}$ ($\rho_{Cl}$).}
\begin{tabular}{ccccccccccc}
\hline\hline
           FeCl$_2$ & a      & d$_{Fe-Cl}$  & d$_{Cl-Cl}$    & $\theta$ & $\mu$    & $\rho_{Fe}$ & $\rho_{Cl}$&  $E_{c}$\\ 
            Phase     & ({\AA})& ({\AA})      &     ({\AA})    &   (deg)  &$(\mu_B)$ & (e)         & (e)        & eV/atom \\
\hline
1T & 3.47 &  2.45 & 2.84  &90.15 & 4 & -1.20 & 0.60& 2.64\\
1H & 3.34 &  2.47 & 3.08  &77.23 & 4 & -0.86 & 0.43& 2.59\\
\hline\hline
\end{tabular}
\end{table}

FeCl$_2$ is not the only possible FeCl$_{n}$ type structure in the monolayer form. 
Indeed, recently it was reported that FeCl$_3$ also exists. \cite{wehenkel} In 
order to investigate the most favorable  chemical composition of the compound, 
we calculated the cohesive energy of  FeCl$_3$. The calculated cohesive 
energy of FeCl$_3$ is 2.4 eV/atom which is smaller than the value for the 1H and 
1T phases of FeCl$_2$. This shows that the 1T phase of FeCl$_2$ is the most 
favorable monolayer structure of the FeCl$_n$ type structures.

\begin{figure}
\includegraphics[width=7.6cm]{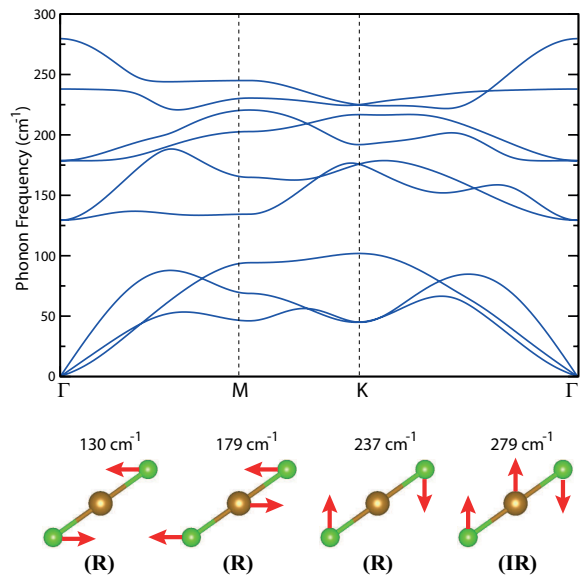}
\caption{\label{fig2}(Color online) Phonon dispersion for the 1T phase of FeCl$_2$. Eigenfrequencies for Raman and Infrared-active modes and corresponding eigenmotions are shown in the lower panel.}
\end{figure}

Phonon dispersion of the single layer  FeCl$_2$ in the 1T phase is shown in Fig. 
\ref{fig2}. Here the dynamical matrix and the vibrational modes 
were calculated using the small-displacement method (SDM)~\cite{alfe}  with 
forces obtained from VASP.  We found that all the phonon branches have real 
eigenfrequencies through all the symmetry points and hence the predicted structure 
of 1T-FeCl$_2$ is dynamically stable.

There is one Fe atom and two Cl atoms within the primitive unit cell of 1T-FeCl$_2$ and its phonon spectrum includes 9 phonon bands, 3 acoustic (transverse acoustic TA, longitudinal acoustic LA, and out-of-plane transverse acoustic ZA) and 6 optical. While LA and TA acoustic branches have characteristics linear dispersion, the frequency of the out of plane ZA mode has a quadratic dispersion in the vicinity of the zone center. Analysis of the lattice dynamics shows that the decomposition of the vibration representation of optical modes (without translational acoustic modes) at the $\Gamma$  point is  $\Gamma_{opt}= 2E' + A{_1}' + 2E'' + A{_2}''$. We also show the optical character and frequency of the phonons for the monolayer FeCl$_2$ in the lower panel of Fig. \ref{fig2}. Note that while the phonon modes that have out-of-plane character ($A{_1}'$ and $A{_2}''$) are singly degenerate, modes with in-plane vibrational character ($E'$ and $E''$) are doubly degenerate. 

In addition a large phonon energy bandgap, which is 130 cm$^{-1}$ at the $\Gamma$ and  30 cm$^{-1}$ at the $K$ point,  is found between the acoustic phonon modes and the optical phonon modes.  Due to the large phonon bandgap, the acoustic vibration is preserved from being interrupted by the optical modes and hence, FeCl$_2$ may have a higher quality factor than a graphene resonator.


\begin{figure}
\includegraphics[width=7.5cm]{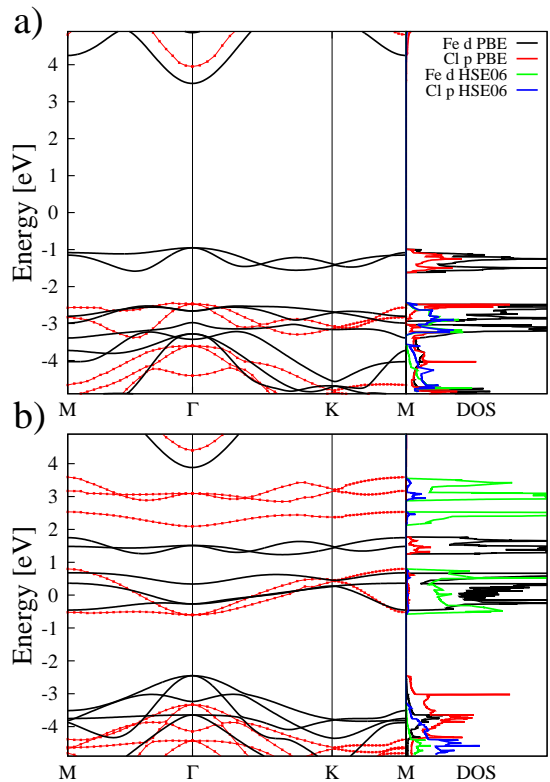}
\caption{\label{fig3}(Color online) Electronic band structure of 2D 
single-layer FeCl$_2$ in the T structure and projected DOS of Fe and Cl atoms 
(a) minority and (b) majority states. The Fermi level of both calculations are set to 0 eV.}
\end{figure}

FeCl$_2$ has a polar character in addition to the covalency of the bonds. Effective charge on cation and anion, $Z_c^*$ and $Z_a^*$, charge transferred from  cation to anion, $\delta\rho=Z_a^*-Z_v$ ($Z_v$ being the valency of the constituent atom)  are calculated using the Bader analysis. In spite of the ambiguities in finding the true effective  charge, the calculated effective charges in Table~\ref{table1} give some idea  about the direction of charge transfer and ionicity of the honeycomb structure.  Bader charge analysis showed that the Fe atom donates 1.20 \textit{e} and each Cl atom receives 0.60 \textit{e} charge in the T structure. In the H structure, however, Fe atom donates \textit{0.86} e charge and  each Cl atom receives \textit{0.43} e charge.     

In Fig.~\ref{fig3} the band structure and the density of states of FeCl$_2$ are 
shown for the minority (Fig.~\ref{fig3}a) and the majority (Fig.~\ref{fig3}b) states. 
The states which are around the Fermi level are mostly from the d-electrons of the Fe 
atom.  The bands which are plotted in black in Fig. \ref{fig3} are 
the results of PBE-GGA calculation. According to PBE-GGA calculations  FeCl$_2$ 
show half-metallic property.  An energy gap around 4.4 eV is observed in the 
electronic structure of the minority states, however there are bands which are 
crossing the Fermi level in the case of the majority states which makes the compound  a 
half metal. On top of PBE-GGA calculations we also performed HSE06 hybrid 
functional calculation to interpret the electronic structure of the compound 
more accurately.  The HSE06 bands are shown in red in the band structure plots. 
As can be seen from the figure the half metallic behavior of  FeCl$_2$ 
remains in the HSE06 calculations and the gap for the  minority electrons 
increased to 6.7 eV. 

\begin{figure}[htbp]
\centering
\includegraphics[width=8.0cm]{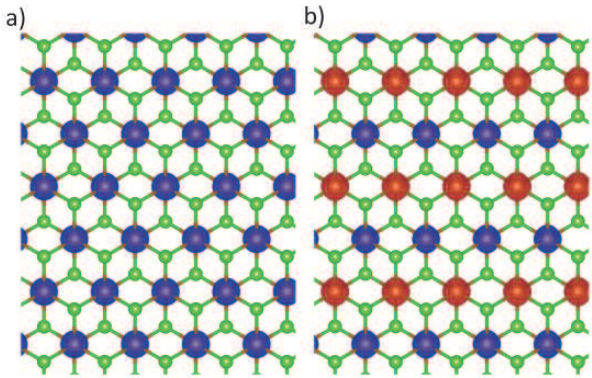}
\caption{(Color online) Spin density plot of (a) ferromagnetic and b) anti-ferromagnetic configurations of FeCl$_2$.  
The blue lobes correspond to the minority and the red ones to the majority spin density. \label{fig4}}
\end{figure}

Ferromagnetically ordered magnetic moments are a crucial requirement for spintronics devices.  To find whether or not ferromagnetic state is energetically favorable, we perform  supercell calculations and compare the energies of ferromagnetic and  anti-ferromagnetic configurations. For FeCl$_2$ we compared the total energies of non-magnetic,  ferromagnetic and  anti-ferromagnetic configurations. The  spin density plots of these calculations are shown in Fig. \ref{fig4}. Since the primitive cell of  FeCl$_2$ has 4 $\mu_B$ magnetic moments, ferromagnetic configuration of  $2\times2$ supercell (Fig. \ref{fig4}a)  carries 16  $\mu_B$. The anti-ferromagnetic configuration of $2\times2$  supercell (Fig. \ref{fig4}b) has 0 $\mu_B$ magnetic moment and 93 meV per primitive cell higher energy than the ferromagnetically ordered one. The non-magnetic configuration of FeCl$_2$ is 1.3 eV per primitive cell higher in energy than the ferromagnetic one. This means that the ferromagnetically ordered state is energetically the most favorable  arrangement of FeCl$_2$. 

Curie temperature is defined as the temperature which separates the paramagnetic 
phase in which the magnetic moments are randomly oriented from the ordered 
ferromagnetic phase.  The Curie temperature of the materials can be approximated 
using the Heisenberg model in which the Hamiltonian can be written as $\hat{H} = 
-\sum_{i,j} J \hat{m}_i.\hat{m}_j$, where $J$ is the Heisenberg exchange 
parameter and $\hat{m}$ is the magnetic moment of each site in $\mu_B$. The 
expression for the Heisenberg exchange parameter for our system is 
$J=(1/12)E_{ex}/2m^{2}$ where $E_{ex}$ is the energy difference between 
ferromagnetic and anti-ferromagnetic configurations and the factor (1/12) 
is due to the double counting of the exchange interaction of 6 nearest neighbor 
atoms in the summation. If we substitute $m=4$ and $E_{xc}=372$ into the 
equation, we get $J=0.97$ meV. There exist several approximations to find $T_{c}$ 
from the $J$ value. Here we will use the mean field approximation which is 
$k_{B}T_{c}=\frac{3}{2}J$.
This approximation leads to a Curie temperature of 17 K. In addition, it is worthwhile to note that one may find different, but in the same order, T$_c$ values using different methodologies \cite{zhou,zhang,sims}. 

To summarize, in this paper we investigated the structural, electronic and magnetic properties of single layers of FeCl$_2$  using first principles calculations. We found that the T phase  is energetically more favorable than the H phase of the compound.  Our vibrational analysis showed that all the phonon branches have real eigenvalues through all symmetry points which indicates the stability of the 1T-FeCl$_2$ structure. The electronic structure of 1T-FeCl$_2$ is investigated using both GGA-PBE and DFT-HSE06 functionals. A very interesting physical property of the compound is revealed with both functionals: FeCl$_2$ has a half-metallic nature. Our supercell calculations showed that the ferromagnetic ordering of the magnetic moments have a lower energy than the anti-ferromagnetic ordered state.  Thus, the  1T-FeCl$_2$ is an intrinsic half-metallic ferromagnet which makes it an attractive  compound for future spintronics devices.
 

This work was supported by the Flemish Science Foundation (FWO-Vl) and
the Methusalem foundation of the Flemish government. Computational
resources were provided by TUBITAK ULAKBIM, High Performance and Grid
Computing Center (TR-Grid e-Infrastructure). H.S. is supported by a
FWO Pegasus Long Marie Curie Fellowship.

\end{document}